%% file: P1-80.tex
\input{aipcheck}

\documentclass[
    ,final            
  ]
  {aipproc}

\layoutstyle{6x9}

\input{myqcircuit}

\newcommand{\Tr}{\operatorname{Tr}}

\newcommand{\floor}[1]{\lfloor#1\rfloor}
\newcommand{\ceil}[1]{\lceil#1\rceil}
\newcommand{\hilb}[1]{\mathcal{#1}}

\begin{document}

\title{Quantum Reading of Unitary Optical Devices}

\classification{03.67.-a, 42.50.Dv, 42.50.Ex}

\keywords{quantum reading, discrimination unitaries, discrimination
  channels}

\author{Michele Dall'Arno}{
  address={Graduate School of Information Science, Nagoya University,
  Nagoya, 464-8601, Japan\\ICFO-Institut de Ciencies Fotoniques,
  E-08860 Castelldefels (Barcelona), Spain\\Quit group, Dipartimento
  di Fisica, via Bassi 6, I-27100 Pavia, Italy}
}

\author{Alessandro Bisio}{
  address={Quit group, Dipartimento di Fisica, via Bassi 6, I-27100
  Pavia, Italy\\Istituto Nazionale di Fisica Nucleare, Gruppo IV, via
  Bassi 6, I-27100 Pavia, Italy}
}

\author{Giacomo Mauro D'Ariano}{
  address={Quit group, Dipartimento di Fisica, via Bassi 6, I-27100
  Pavia, Italy\\Istituto Nazionale di Fisica Nucleare, Gruppo IV, via
  Bassi 6, I-27100 Pavia, Italy}
}

\begin{abstract}
  We address the problem of quantum reading of optical memories,
  namely the retrieving of classical information stored in the optical
  properties of a media with minimum energy. We present optimal
  strategies for ambiguous and unambiguous quantum reading of unitary
  optical memories, namely when one's task is to minimize the
  probability of errors in the retrieved information and when perfect
  retrieving of information is achieved probabilistically,
  respectively. A comparison of the optimal strategy with coherent
  probes and homodyne detection shows that the former saves orders of
  magnitude of energy when achieving the same
  performances. Experimental proposals for quantum reading which are
  feasible with present quantum optical technology are reported.
\end{abstract}

\maketitle


In the engineering of optical memories (such as CDs or DVDs), a
tradeoff among several parameters must be taken into account. High
precision in the retrieving of information is surely an indefeasible
assumption, but also energy requirements, size and weight can play a
very relevant role for applications. Clearly using a low energetic
radiation to read information reduces the heating of the physical bit,
thus allowing for smaller implementation of the bit itself. In the
problem of {\em quantum reading}~\cite{Pir11, BDD11, Nai11, PLGMB11,
Hir11, DBDMJD12, SLMBP12, NYGSP12, PUR12, DBD13, Dal13} of optical
memories one's task is to exploit the quantum properties of light to
retrieve some classical digital information stored in the optical
properties of a given media with minimum energy. We focus on the case
where information is encoded into linear and energy-preserving unitary
optical devices~\cite{BDD11,DBDMJD12,DBD13}, while most of the
previous literature focused on the case of non unitary - e.g. lossy -
devices.

In this hypothesis two different scenarios can be distinguished. A
possibility is the on-the-fly retrieving of information
(e.g. multimedia streaming), where one requires that the reading
operation is performed fast - namely, only once, but a modest amount
of errors in the retrieved information is tolerable. In this context,
denoted as ambiguous quantum reading, the relevant figure of merit is
the probability $P_e$ to have an error in the retrieved
information. On the other hand for highly reliable technology, perfect
retrieving of information is an issue. Then, unambiguous quantum
reading, where one allows for an inconclusive outcome (while, in case
of conclusive outcome, the probability of error is zero) becomes
essential. Here, the relevant figure of merit is clearly the failure
probability $P_f$ of getting an inconclusive outcome.

We discuss~\cite{BDD11, DBDMJD12, DBD13} optimal strategies for both
scenarios, which exploit fundamental properties of quantum theory such
as entanglement, allowing for the ambiguous (unambiguous)
discrimination of linear and energy-preserving unitary devices with
probability of error $P_e$ (probability of failure $P_f$) under any
given threshold, while minimizing the energy requirement. The most
general strategy for performing quantum reading consists in preparing
a bipartite probe $\rho$ (we allow for an ancillary mode), applying
locally the unknown device and performing a bipartite POVM $\Pi$ on
the output state, namely
\begin{align*}
  \begin{aligned}
    \Qcircuit @C=0.7em @R=1em { \multiprepareC{1}{\rho} &
      \ustick{\hilb{H}} \qw & \gate{U_x} & \qw &
      \ghost{\Pi}\\ \pureghost{\rho} & \ustick{\hilb{K}} \qw & \qw &
      \qw & \multimeasureD{-1}{\Pi} }
  \end{aligned}.
\end{align*}

Since the optimal POVMs and the corresponding error (failure)
probabilities for ambiguous (unambiguous) discrimination of two states
are well known, the problem of quantum reading can be formulated as an
optimization over probe only. For any set of two optical devices
$\{U_1, U_2\}$ and any threshold $q$ in the probability of error
(failure), find the minimum energy probe $\rho^*$ that allows to
ambiguously (unambiguously) discriminate between $U_0$ and $U_1$ (with
equal priors) with probability of error $P_e$ (probability of failure
$P_f$) not larger than $q$, namely
\begin{align*}
  \rho^* = \arg \min_{\rho \textrm{ s.t. } P(\rho, U_1, U_2) \leq q}
  E(\rho).
\end{align*}
where $P(\rho, U_0, U_1)$ can either be given by $P_e$ or $P_f$ and
$E(\rho) := \Tr[\rho N]$ is the energy (expectation value of number
operator $N$) of probe $\rho$. Notice that the discrimination of
devices $\{U_1, U_2\}$ can be easily recasted to that of devices $\{I,
U := U_1^{-1}U_2\}$. In the following we will write Fock states
$\ket{n,m}$ in the basis where $U$ is diagonal.

We are now ready to present our main results~\cite{BDD11, DBDMJD12,
DBD13}. It is possible to prove that without loss of generality the
optimal probe $\rho^*$ for (ambiguous or unambiguous) quantum reading
can be taken pure and no ancillary modes are required. For the quantum
reading of beamsplitters, the optimal probe $\rho^*$ is given by a
superposition of a NOON state and the vacuum, namely
\begin{align*}
  \ket{\psi^*} = \alpha \frac{\ket{0,n^*} + \ket{n^*,0}}{\sqrt{2}} +
  \sqrt{1-\alpha^2} \ket{0,0},
\end{align*}
where $|\alpha| = \sqrt{\frac{1-K}{1-\cos{\delta n^*}}}$, $n^* =
\arg\min_{\floor{x^*},\ceil{x^*}} E(\psi^*)$, $x^* = \min[x \ge 0 |
  \delta x = \tan(\delta x / 2)]$, $K = \sqrt{4q(1-q)}$ for ambiguous
quantum reading while $K = q$ for unambiguous quantum reading, and the
eigenvalues of the scattering matrix of the beamsplitter are $e^{\pm
i \delta}$.

\begin{figure}
  \label{fig:tradeoff}

  \includegraphics[height=.3\textheight]{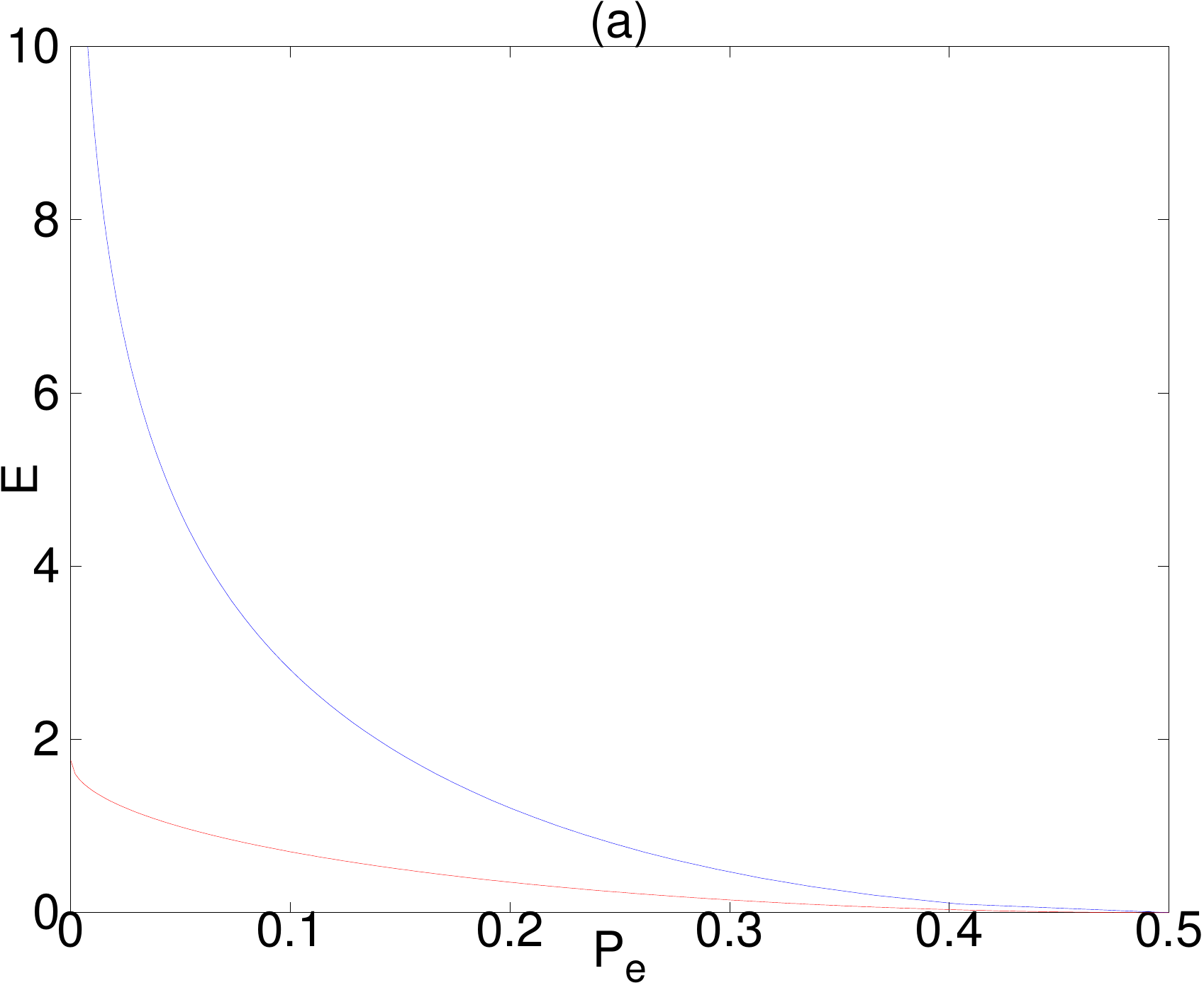}

  \includegraphics[height=.3\textheight]{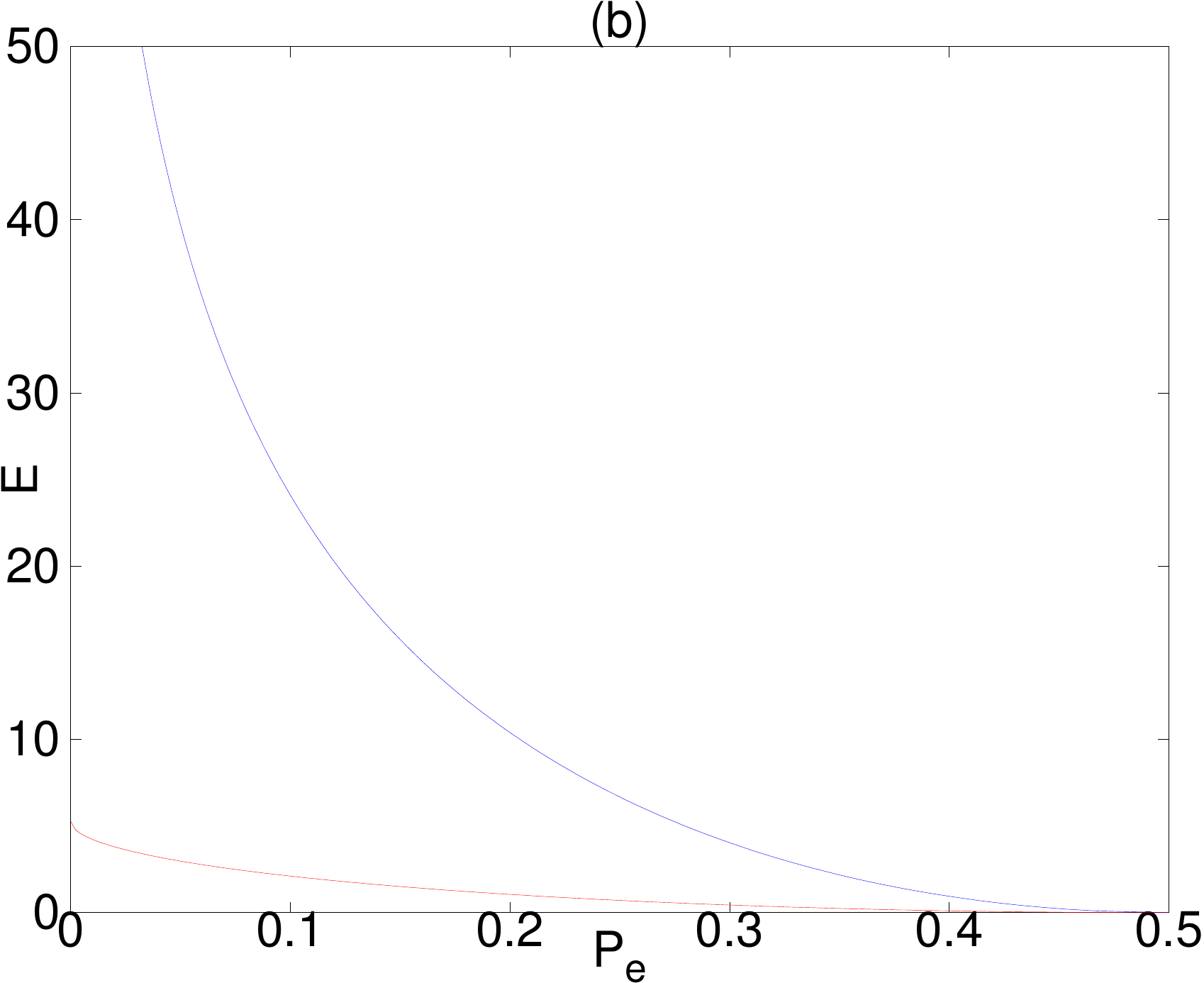}

  \caption{(Color online) Optimal tradeoff between the energy $E$ and
  the probability of error $P_e$ in the discrimination of $I$ and
  $U=\exp(i (\delta a_1^\dagger a_1 - \delta a_2^\dagger a_2))$
  ($\delta=\pi/4$ in (a) and $\delta=\pi/12$ in (b)). The upper line
  represents the discrimination with coherent probe and homodyne
  detection, while the lower line represents the optimal
  discrimination.}
\end{figure}

A comparison of the optimal strategy for ambiguous quantum reading
with the optimal coherent strategy (using coherent probes and homodyne
detection), reminiscent of the one implemented in common CD readers,
is provided in Fig.~\ref{fig:tradeoff}. The Figure clarifies that the
former strategy saves orders of magnitude of energy, moreover allowing
for perfect discrimination with finite energy.
We underline that experimental proposals~\cite{DBDMJD12} for ambiguous
and unambiguous optimal quantum reading were provided for the
single-photon case - namely, $n^*=1$. They are feasible with present
quantum optical technology, in terms of single-photon source, linear
optics and photodetectors.

In this work we addressed the problem of quantum reading of linear and
energy-preserving unitary optical memories. We showed that the probe
can be taken pure and no ancillary modes are needed. For quantum
reading of beamsplitters, we presented optimal strategies for
ambiguous (unambiguous) quantum reading, where the probe is given by a
superposition of a NOON state and the vacuum. We compared the optimal
quantum strategy with a coherent strategy, showing that the former
saves orders of magnitude of energy when compared with the latter, and
we discussed experimental feasibility.


\begin{theacknowledgments}
  This work was supported by JSPS (Japan Society for the Promotion of
  Science) Grant-in-Aid for JSPS Fellows No. 24-0219, Spanish project
  FIS2010-14830, Italian project PRIN 2008, and European projects
  COQUIT and Q-Essence.
\end{theacknowledgments}



\bibliographystyle{aipproc}   


\IfFileExists{\jobname.bbl}{}
 {\typeout{}
  \typeout{******************************************}
  \typeout{** Please run "bibtex \jobname" to optain}
  \typeout{** the bibliography and then re-run LaTeX}
  \typeout{** twice to fix the references!}
  \typeout{******************************************}
  \typeout{}
 }


\end{document}

%% file: myqcircuit.tex
\usepackage[matrix,frame,arrow]{xy}
\usepackage{amsmath}

\newcommand{\ket}[1]{\left\vert{#1}\right\rangle}
\newcommand{\qw}[1][-1]{\ar @{-} [0,#1]}

\newcommand{\gate}[1]{*{\xy *+<.6em>{#1};p\save+LU;+RU **\dir{-}\restore\save+RU;+RD **\dir{-}\restore\save+RD;+LD **\dir{-}\restore\POS+LD;+LU **\dir{-}\endxy} \qw}

\newcommand{\multimeasureD}[2]{*+<1em,.9em>{\hphantom{#2}}\save[0,0].[#1,0];p\save !C *{#2},p+LU+<0em,0em>;+RU+<-.8em,0em> **\dir{-}\restore\save +LD;+LU **\dir{-}\restore\save +LD;+RD-<.8em,0em> **\dir{-} \restore\save +RD+<0em,.8em>;+RU-<0em,.8em> **\dir{-} \restore \POS !UR*!UR{\cir<.9em>{r_d}};!DR*!DR{\cir<.9em>{d_l}}\restore \qw}

\newcommand{\ghost}[1]{*+<1em,.9em>{\hphantom{#1}} \qw}

\newcommand{\ustick}[1]{*!D!<0em,-.5em>=<0em>{#1}}

\newcommand{\Qcircuit}[1][0em]{\xymatrix @*=<#1>}

\newcommand{\pureghost}[1]{*+<1em,.9em>{\hphantom{#1}}}
\newcommand{\multiprepareC}[2]{*+<1em,.9em>{\hphantom{#2}}\save[0,0].[#1,0];p\save !C
  *{#2},p+RU+<0em,0em>;+LU+<+.8em,0em> **\dir{-}\restore\save +RD;+RU **\dir{-}\restore\save
  +RD;+LD+<.8em,0em> **\dir{-} \restore\save +LD+<0em,.8em>;+LU-<0em,.8em> **\dir{-} \restore \POS
  !UL*!UL{\cir<.9em>{u_r}};!DL*!DL{\cir<.9em>{l_u}}\restore}


%% file: P1-80.bbl
\begin{thebibliography}{}

\bibitem{Pir11} S. Pirandola, Phys. Rev. Lett. {\bf 106}, 090504
  (2011).

\bibitem{BDD11} A. Bisio, M. Dall'Arno, and G. M. D'Ariano,
  Phys. Rev. A {\bf 84}, 012310 (2011).

\bibitem{Nai11} R. Nair, Phys. Rev. A {\bf 84}, 032312 (2011).

\bibitem{PLGMB11} S. Pirandola, C. Lupo, V. Giovannetti, S. Mancini,
  and S. L. Braunstein, New J. Phys. {\bf 13}, 113012 (2011).

\bibitem{Hir11} O. Hirota, arXiv:1108.4163.

\bibitem{DBDMJD12} M. Dall'Arno, A. Bisio, G. M. D'Ariano, M. Mikova,
  M. Jezek, and M. Dusek, Phys. Rev. A {\bf 85}, 012308 (2012).

\bibitem{SLMBP12} G. Spedalieri, C. Lupo, S. Mancini,
  S. L. Braunstein, and S. Pirandola, Phys. Rev. A {\bf 86}, 012315
  (2012).

\bibitem{NYGSP12} R. Nair, B. J. Yen, S. Guha, J. H. Shapiro,
  S. Pirandola, Phys. Rev. A {\bf 86}, 022306 (2012).

\bibitem{PUR12} J. Prabhu Tej, A. R. Usha Devi, A. K. Rajagopal,
  arXiv:1210.0791.

\bibitem{DBD13} M. Dall'Arno, A. Bisio, and G. M. D'Ariano, J. Phys.:
  Conf. Ser. {\bf 414}, 012038 (2013).

\bibitem{Dal13} M. Dall'Arno, arXiv:1302.1624.

\end{thebibliography}
